\documentclass[sigconf]{acmart}

\usepackage{textcomp}
\usepackage{stfloats}
\usepackage{url}
\usepackage{verbatim}
\usepackage{amsthm}
\usepackage{subfigure}
\usepackage{makecell}
\usepackage{multirow}
\usepackage{enumitem}
\usepackage{subcaption}

\usepackage{color, colortbl}

\definecolor{Gray}{rgb}{0.9, 0.9, 0.9}

\newcommand{\gray}{\cellcolor{Gray}}

\usepackage[capitalize,noabbrev]{cleveref}

\usepackage{booktabs}
\usepackage{fontawesome}

\usepackage[ruled,linesnumbered]{algorithm2e}

\usepackage[inkscapelatex=false]{svg}

\copyrightyear{2026}
\acmYear{2026}
\setcopyright{cc}
\setcctype{by}
\acmConference[KDD 2026] {Proceedings of the 32nd ACM SIGKDD Conference on Knowledge Discovery and Data Mining V.1}{August 9--13, 2025}{Jeju Island, Republic of Korea.}
\acmBooktitle{Proceedings of the 32nd ACM SIGKDD Conference on Knowledge Discovery and Data Mining V.1 (KDD 2026), August 9--13, 2025, Jeju Island, Republic of Korea}
\acmISBN{979-8-4007-2258-5/2026/08}
\acmDOI{10.1145/3770854.3780200}

\begin{document}

\title[VI-MMRec]{VI-MMRec: Similarity-Aware Training Cost-free Virtual User-Item Interactions for Multimodal Recommendation}

\author{Jinfeng Xu}
\email{jinfeng@connect.hku.hk}
\affiliation{%
  \institution{The University of Hong Kong}
  \city{HongKong SAR}
  \country{China}}

\author{Zheyu Chen}
\email{zheyu.chen@bit.edu.cn}
\affiliation{%
  \institution{Beijing Institute of Technology}
  \city{Beijing}
  \country{China}}

\author{Shuo Yang}
\email{shuoyang.ee@gmail.com}
\affiliation{%
  \institution{The University of Hong Kong}
  \city{HongKong SAR}
  \country{China}}

\author{Jinze Li}
\email{lijinze-hku@connect.hku.hk}
\affiliation{%
  \institution{The University of Hong Kong}
  \city{HongKong SAR}
  \country{China}}

\author{Zitong Wan}
\email{zitong.wan@ucdconnect.ie}
\affiliation{
  \institution{University College Dublin}
  \city{Dublin}
  \country{Ireland}}

\author{Hewei Wang}
\email{heweiw@alumni.cmu.edu}
\affiliation{%
    \institution{Carnegie Mellon University}  
    \city{Pittsburgh, PA}   
    \country{USA}}

\author{Weijie Liu}
\email{liuwj0817@connect.hku.hk}
\affiliation{
  \institution{The University of Hong Kong}
  \city{HongKong SAR}
  \country{China}}

\author{Yijie Li}
\email{yijieli@andrew.cmu.edu}
\affiliation{%
  \institution{Carnegie Mellon University}
  \city{Pittsburgh, PA}
  \country{USA}}

\author{Edith C. H. Ngai}
\authornote{Corresponding authors}
\email{chngai@eee.hku.hk}
\affiliation{%
  \institution{The University of Hong Kong}
  \city{HongKong SAR}
  \country{China}}

\renewcommand{\shortauthors}{Jinfeng Xu et al.}


 
\begin{abstract}
Although existing multimodal recommendation models have shown promising performance, their effectiveness continues to be limited by the pervasive data sparsity problem. This problem arises because users typically interact with only a small subset of available items, leading existing models to arbitrarily treat unobserved items as negative samples. To this end, we propose VI-MMRec, a model-agnostic and training cost-free framework that enriches sparse user-item interactions via similarity-aware virtual user-item interactions. These virtual interactions are constructed based on modality-specific feature similarities of user-interacted items. Specifically, VI-MMRec introduces two different strategies: (1) Overlay, which independently aggregates modality-specific similarities to preserve modality-specific user preferences, and (2) Synergistic, which holistically fuses cross-modal similarities to capture complementary user preferences. To ensure high-quality augmentation, we design a statistically informed weight allocation mechanism that adaptively assigns weights to virtual user-item interactions based on dataset-specific modality relevance. As a plug-and-play framework, VI-MMRec seamlessly integrates with existing models to enhance their performance without modifying their core architecture. Its flexibility allows it to be easily incorporated into various existing models, maximizing performance with minimal implementation effort. Moreover, VI-MMRec introduces no additional overhead during training, making it significantly advantageous for practical deployment. Comprehensive experiments conducted on six real-world datasets using seven state-of-the-art multimodal recommendation models validate the effectiveness of our VI-MMRec.
\end{abstract}

\ccsdesc[500]{Information systems~Recommender systems;}

\keywords{Mutimodal, Recommendation, Virtual Interactions}

\maketitle

\section{Introduction}
The rapid growth of the internet has resulted in an information explosion, making recommender systems essential for efficiently navigating and filtering vast amounts of data \cite{xu2025nlgcl,chen2025squeeze}. Traditional recommender systems primarily model user preferences based on historical user-item interactions \cite{he2020lightgcn,xu2024fourierkan,xu2024aligngroup}. However, their performance is often constrained by the data sparsity problem, as these systems rely solely on a limited number of historical interactions to infer user preferences and item properties. Recommendation systems have evolved significantly by harnessing multimodal information to address the inherent challenge of data sparsity. Previous works, such as VBPR \cite{he2016vbpr}, pioneered the integration of visual features into matrix factorization frameworks to enhance collaborative filtering. Subsequent advancements expanded this paradigm by incorporating both visual and textual modalities \cite{chen2019personalized,liu2019user,yu2023multi,xu2024mentor, chen2025don}, leveraging richer item representations to improve recommendation performance. The emergence of graph-based methods further propelled progress: MMGCN \cite{wei2019mmgcn} introduced modality-specific graph convolutional networks (GCNs), while DualGNN \cite{wang2021dualgnn} and LATTICE \cite{zhang2021mining} refined user-user and item-item graph structures to model preference commonalities. Building on these foundations, FREEDOM \cite{zhou2023tale} enhanced graph stability through semantic graph freezing and noise mitigation. Recent innovations have focused on self-supervised learning and inter-modal alignment to reduce reliance on labeled data. MMSSL \cite{wei2023multi} and MICRO \cite{zhang2022latent} employed contrastive learning to align multimodal signals with collaborative signals, while BM3 \cite{zhou2023bootstrap} explicitly modeled inter-modal relationships for improved fusion. MENTOR \cite{xu2024mentor} introduced multi-level self-supervised tasks to preserve interaction signals during modality alignment. More recently, LGMRec \cite{guo2024lgmrec} and DiffMM \cite{jiang2024diffmm} adopt hyper-graph architecture and diffusion models to address global-local relations and modality noise reduction, respectively.

Despite these advancements, data sparsity remains a critical bottleneck in real-world scenarios. Because users interact with only a small fraction of items \cite{liu2024multimodal}, leading existing models to arbitrarily treat unobserved items as negative samples—a practice that overlooks latent user preferences. To bridge this gap, we propose VI-MMRec, a model-agnostic framework that mitigates data sparsity by constructing similarity-aware virtual user-item interactions derived from modality-specific feature similarities of interacted items. VI-MMRec introduces two different strategies: 
\begin{itemize}[leftmargin=*]
    \item \textbf{Overlay:} Independently aggregates modality-specific similarities to preserve modality-specific user preferences.
    \item \textbf{Synergistic:} Holistically fuses cross-modal similarities to capture complementary user preferences.
\end{itemize}

To ensure high-quality augmentation, VI-MMRec incorporates a statistically informed weight allocation mechanism that adaptively assigns weights to virtual interactions based on dataset-specific modality relevance, avoiding heuristic or arbitrary weighting. As a plug-and-play framework, VI-MMRec seamlessly integrates with all existing models to enhance their performance without modifying their core architecture. Its flexibility allows it to be easily incorporated into various existing models, maximizing performance with minimal implementation effort. We emphasize that virtual user-item interactions are constructed based on raw modality features and are completely frozen during training, which directly reconstructs user-item interactions without altering the shape of the original user-item interaction matrix. Thus, VI-MMRec introduces no additional overhead during training, making it significantly advantageous for practical deployment. Comprehensive experiments on six real-world datasets using seven advanced multimodal recommendation models demonstrated the effectiveness of our VI-MMRec. In summary, our contributions include:
\begin{itemize}[leftmargin=*]
    \item We propose a novel and plug-and-play framework for augmenting real user-item interactions via similarity-aware virtual user-item interactions, which can be seamlessly adopted in all existing multimodal recommendation models.
    \item We propose two strategies for constructing virtual user-item interactions and an adaptive weighting mechanism grounded in statistical modality relevance. 
    \item VI-MMRec introduces no additional overhead during training, making it significantly advantageous for practical deployment.
    \item Comprehensive experiments on six real-world datasets using seven advanced multimodal recommendation models validate the effectiveness of our VI-MMRec. 
\end{itemize}

\section{Preliminary}
Let $\mathcal{U}=\{u_1,...,u_{|\mathcal{U}|}\}$ denotes user set and $\mathcal{I}=\{i_1,...,i_{|\mathcal{I}|}\}$ denotes item set. We conceptualize the user-item graph $\mathcal{G} = (\mathcal{U}, \mathcal{I}, \mathcal{E})$, where $\mathcal{U}, \mathcal{I}$ serve as the graph vertices, and $\mathcal{E}$ denotes the edge set. In multimodal scenarios, each item contains multiple features. We introduce modality-specific item embedding $i^m$ for each item $i$ belonging to the set of modalities $\mathcal{M}$. The user-item interaction matrix is denoted as $\mathcal{R} \in \{0,1\}^{|\mathcal{U}| \times |\mathcal{I}|}$. The entire user-item interaction matrix can be divided into an observed user-item interaction matrix $\mathcal{R}^{+} \in \{\mathcal{R}_{u,i}|u \in \mathcal{U},i \in \mathcal{I},\mathcal{R}_{u,i} = 1\}$ and an unobserved user-item interaction matrix $\mathcal{R}^{-} \in \{\mathcal{R}_{u,i}|u \in \mathcal{U},i \in \mathcal{I},\mathcal{R}_{u,i} = 0\}$. For each user-item pair $(u,i)$ that satisfies $\mathcal{R}_{u,i} = 1$, there exists bidirectional edges $(u,i) \in \mathcal{E}$ and $(i,u) \in \mathcal{E}$. 

We random initialize $\mathbf{E}_{u^m} \in \mathbb{R}^{d_m \times |\mathcal{U}|}$ to represent user embedding with modality $m$. $\mathbf{E}_{i^m} \in \mathbb{R}^{d_m \times|\mathcal{I}|}$ represents item initialized embedding with modality $m$, which extracted by pre-trained encoders. Here $d_m$ represents the hidden dimensionality. Based on the user-item graph $\mathcal{G}$, most existing works \cite{xu2024mentor,zhou2023tale,wei2019mmgcn} adopt aggregation to enhance user/item embeddings for extracting high-order user-item collaborative signals. Take the most widely-used GNN backbone LightGCN \cite{he2020lightgcn} as an example, the embeddings for user $u$ and item $i$ in the $l$-th layer are:
\begin{equation}
\label{eq:1}
\mathbf{e}_{u_{m}}^{(l)}=\frac{1}{d_u}\sum_{j | (u,j) \in \mathcal{E}} \frac{1}{d_j}\mathbf{e}_{j_{m}}^{(l-1)},  \quad \mathbf{e}_{i_{m}}^{(l)}=\frac{1}{d_i}\sum_{v |(i,v)\in\mathcal{E}} \frac{1}{d_v}\mathbf{e}_{v_{m}}^{(l-1)},
\end{equation}
where $d_{*}$ denotes degree of node. $\mathbf{e}_{*}^{(l)}$ represents node embedding in $l$-th layer. After $L$ layers of neighbor aggregation, the final representations of modality $m$ for user $u$ and item $i$ are expressed as:
\begin{equation}
\label{eq:2}
    \mathbf{\bar{e}}_{u_{m}} = \sum_{l=0}^{L} \mathbf{e}^{l}_{u_{m}}, \quad \mathbf{\bar{e}}_{i_{m}} = \sum_{l=0}^{L} \mathbf{e}^{l}_{i_{m}}.
\end{equation}

For model optimization, most previous works adopt the BPR \cite{rendle2012bpr} loss to train the model:
\begin{equation}
\label{eq:bpr}
    \mathcal{L}_{bpr} = \sum_{(u, i^{+}, i^{-}) \in \mathcal{D}} - \log(\sigma(\hat{y}_{u,i^{+}} - \hat{y}_{u,i^{-}})),
\end{equation}
where $(u, i^{+}) \in \mathcal{R}^{+}$ and $(u, i^{-}) \in \mathcal{R}^{-}$ for triplet $(u, i^{+}, i^{-}) \in \mathcal{D}$. $\sigma(\cdot)$ denotes activation function. The predicted user-item relation score can be calculated by $\hat{y}_{u,i}=\sum_{m \in \mathcal{M}}(\bar{\mathbf{e}}_{u_{m}}^{\top} \bar{\mathbf{e}}_{i_{m}})$. 

Although the above multimodal recommendation paradigms have demonstrated promising performance, their effectiveness is limited by the problem of data sparsity. Data sparsity arises because users typically interact with only a small subset of available items in real-world scenarios \cite{liu2024multimodal, xu2025survey}. Consequently, certain items that users might be interested in but have no interaction records are often arbitrarily categorized as negative samples by existing models \cite{wang2023missrec,zhou2023tale}.

An underexplored but strongly real-world intuitive idea is to build similarity-aware virtual interactions to augment user-item interaction density. This idea relies on strong real-world motivation \cite{zhou2023comprehensive,liu2024multimodal,jiang2024diffmm,guo2024lgmrec}: users have relatively fixed modality preferences, e.g., a user likes a specific cartoon pattern, which makes items with similar visual modality representations to the items purchased by the user more likely to be the items that the user is interested in but missed. To verify the effectiveness of this idea, we conducted an empirical study in Section~\ref{sec:investigation} to explore the relations between user-interacted items and items with high modal similarity to these interacted items across various widely used real-world datasets. This investigation strongly supports our plug-and-play framework proposed in Section~\ref{sec:methodology}, which effectively builds similarity-aware virtual user-item interactions to mitigate the data sparsity problem.

\section{Investigation}
\label{sec:investigation}
In this section, we empirically verify the feasibility of constructing similarity-aware virtual interactions by exploring the relations between user-interacted items and items with high modality similarity to these interacted items across various widely used real-world datasets. Specifically, we explore the following aspects:
\begin{itemize}[leftmargin=*]
    \item We calculate the overlap rate between all similarity-aware virtual interactions and real user-item interactions to verify whether modality similarity is correlated with user preferences.
    \item We calculate the overlap rate between the overlapping parts of modality-specific virtual interactions from different modalities and real user-item interactions to verify whether multiple modal similarities have synergistic effects on inferring user preferences.
    \item We calculate the overlap rate within similarity-aware virtual interactions for each modality to verify whether users always have relatively fixed modality preferences.
\end{itemize}

\subsection{Settings}
\label{subsec:investigation settings}
In this investigation, all implementations are built upon MMRec \cite{zhou2023mmrecsm}, the most widely adopted library in prior multimodal recommendation research \cite{zhou2023tale,xu2024mentor,jiang2024diffmm,zhou2023bootstrap}, specifically designed for recommendation tasks involving visual and textual modalities. We utilize popular datasets—Baby, Sports, Clothing, Pet, and Office—widely used in the multimodal recommendation field \cite{mcauley2015image, xu2025survey}. To further evaluate the performance of our pluggable framework proposed in Section~\ref{sec:methodology}, we conduct experiments on the TikTok dataset \footnote{\href{https://www.tiktok.com/}{https://www.tiktok.com/}}, which incorporates an additional audio modality, as detailed in Section~\ref{subsec:RQ1-2}.

\subsection{Overlap Rate Calculation}

For each user-item interaction $(u,i)$ that satisfies $\mathcal{R}_{u,i} \in \mathcal{R}^{+}$, we can get top-$k$ similar items for item $i$ based on item modality-specific features. Specifically, the similarity between item pairs can be calculated as follows:
\begin{equation}
    \mathcal{S}^m_{i,j}=\frac{(\mathbf{e}_{i^m})^{\top} \mathbf{e}_{j^m}}{\|\mathbf{e}_{i^m}\|\|\mathbf{e}_{j^m}\|},
\end{equation}
where $\mathbf{e}_{i^{m}}$ and $\mathbf{e}_{j^{m}}$ denote the row $i$ and row $j$ of item embedding $\mathbf{E}_{i^m}$ with modality $m$, respectively.
We build modality-specific item sets $\mathcal{I}^{m}_{u,i}$ to retain top-$k$ neighbors with the highest similarity, formally: each $j \in \mathcal{I}^{m}_{u,i}$ satisfies $\mathcal{S}^m_{i,j} \in \text { top-} k(\mathcal{S}^m_{i,p}|p \in \mathcal{I},p \neq i)$. 

Furthermore, we can build $|\mathcal{M}|$ similarity-aware virtual user-item interaction matrices $\mathcal{R}^{m}$ based on all modality-specific item sets. Formally, each similarity-aware virtual user-item interaction $\mathcal{R}^{m}_{u,j}$ satisfies $j \in \mathcal{I}^{m}_{u,i}$, where $\mathcal{R}_{u,i} \in \mathcal{R}^{+}$. Moreover, we construct a V-T synergistic virtual user-item interaction matrix $\mathcal{R}^{vt}$ based on visual and textual modalities\footnote{Based on our settings mentioned in Section~\ref{subsec:investigation settings}, our investigation only contains visual and textual modalities.}. Formally, each virtual user-item interaction $\mathcal{R}^{vt}_{u,j}$ satisfies $j \in \mathcal{I}^{vt}_{u,i}$, where $\mathcal{S}^{vt}_{i,j} \in \text { top-} k(\mathcal{S}^v_{i,p} + \mathcal{S}^t_{i,p}|p \in \mathcal{I},p \neq i)$.

The overlap rate $\mathcal{O}^{m}_{real}$ between the similarity-aware virtual user-item interaction matrices $\mathcal{R}^{m}$ and the real user-item interaction matrix $\mathcal{R}^{+}$ is defined as the proportion of interactions in the similarity-aware virtual user-item interaction matrices that are included in the real user-item interaction matrix $\mathcal{R}^{+}$, relative to the total number of interactions in the similarity-aware virtual user-item interaction matrices. Formally, the overlap rate can be calculated as:
\begin{equation}
\label{eq:modality overlap}
    \mathcal{O}^{m}_{real} = \frac{|\mathcal{R}^{m} \cap \mathcal{R}^{+}|}{|\mathcal{R}^{+}|},
\end{equation}
where higher $\mathcal{O}^{m}_{real}$ indicates that the items that interacted with users exhibit high similarity in modality $m$, implying that the features of modality $m$ can effectively capture and reflect user preferences. Furthermore, the overlap rate $\mathcal{O}^{vt}_{real}$ between the V-T synergistic virtual user-item interaction matrix $\mathcal{R}^{vt}$ and the real user-item interaction matrix $\mathcal{R}^{+}$ can be calculated as:
\begin{equation}
\label{eq:synergistic overlap}
    \mathcal{O}^{vt}_{real} = \frac{|\mathcal{R}^{vt} \cap \mathcal{R}^{+}|}{|\mathcal{R}^{+}|},
\end{equation}
where higher $\mathcal{O}^{vt}_{real}$ indicates that the items that interacted with users exhibit high similarity in both visual and textual modalities, implying that the features of visual and textual modalities can effectively capture and reflect user preferences. Moreover, it also illustrates that multiple modality similarities have synergistic effects on inferring user preferences. Furthermore, the overlap rate $\mathcal{O}^{m}_{self}$ within similarity-aware virtual interactions for each modality is defined as the ratio between the ideal total interaction count $k*|\mathcal{R}^{+}|$ (under the assumption that all $k$ similarity-aware virtual user-item interactions generated from each real user-item interactions are mutually non-repeating) and the actual total interaction count $|\mathcal{R}^{m}|$ in the similarity-aware virtual user-item interaction matrix. Formally, 
\begin{equation}
    \mathcal{O}^{m}_{self} = \frac{k*|\mathcal{R}^{+}|}{|\mathcal{R}^{m}|},
\end{equation}
where higher $\mathcal{O}^{m}_{self}$ implies that users always have relatively fixed modality preferences.

\subsection{Statistical Analysis}
We conduct a statistical analysis on five widely-used public datasets following the settings mentioned in Section~\ref{subsec:investigation settings}. As shown in Table~\ref{tab:investigation}, we provide statistical analysis for various $k$ values within $\{5, 10, 20\}$.

\begin{table}[!t]
    \centering
\caption{Statistics analysis.}
\vskip -0.15in
\label{tab:investigation}
\resizebox{\linewidth}{!}{
    \begin{tabular}{c|c|ccccc}
     \toprule\toprule
         Dataset& $k$ &  Baby&  Sports&  Clothing&  Pet&  Office\\
         \midrule
         \multicolumn{2}{c|}{\# Users}& 19,445& 35,598& 39,387& 19,856& 4,905\\
         \multicolumn{2}{c|}{\# Items}& 7,050& 18,357& 23,033& 8,510& 2,420\\
         \multicolumn{2}{c|}{\# Interactions}& 160,792& 296,337& 278,677& 157,836& 53,258\\
         \multicolumn{2}{c|}{\# Sparsity}& 99.88\%& 99.95\%& 99.97\%& 99.91\%& 99.55\%\\
         \multicolumn{2}{c|}{\# Avg. Inter./U}& 8.27& 8.32& 7.08& 7.95& 10.86\\
         \multicolumn{2}{c|}{\# Avg. Inter./I}& 22.81& 16.14& 12.10& 18.55& 22.10\\
         \multicolumn{2}{c|}{\# Matrix Size}& 137.09M& 653.47M& 907.20M& 168.97M& 11.87M\\
         \midrule\midrule
         \multirow{3}{*}{$\mathcal{O}^{v}_{real}$}& 5& 2.60\%& 2.94\%& 3.27\%& 6.16\%& 10.46\%\\
         & 10& 3.87\%& 4.16\%& 4.76\%& 8.00\%& 15.07\%\\
         & 20& 5.83\%& 5.76\%& 6.83\%& 10.16\%& 20.70\%\\\midrule
         \multirow{3}{*}{$\mathcal{O}^{t}_{real}$}& 5& 5.41\%& 7.08\%& 7.40\%& 11.73\%& 14.58\%\\
         & 10& 8.27\%& 10.42\%& 10.47\%& 16.35\%& 21.88\%\\
         & 20& 12.07\%& 14.63\%& 14.25\%& 21.63\%& 30.80\%\\\midrule
         \multirow{3}{*}{$\mathcal{O}^{vt}_{real}$}& 5& 6.74\%& 8.31\%& 9.06\%& 13.70\%& 19.23\%\\
         & 10& 10.16\%& 11.99\%& 12.63\%& 18.47\%& 27.48\%\\
         & 20& 14.86\%& 16.70\%& 17.03\%& 24.16\%& 37.48\%\\\midrule
         \multirow{3}{*}{$\mathcal{O}_{avg}$}& 5& 0.59\%& 0.23\%& 0.15\%& 0.47\%& 2.24\%\\
         & 10& 1.17\%& 0.45\%& 0.31\%& 0.93\%& 4.49\%\\
         & 20& 2.35\%& 0.91\%& 0.61\%& 1.87\%& 8.97\%\\\midrule\midrule
         \multirow{3}{*}{$\mathcal{O}^{v}_{self}$}& 5& 138.28\%& 137.73\%& 143.22\%& 141.68\%& 140.22\%\\
         & 10& 141.03\%& 138.88\%& 144.30\%& 145.05\%& 148.06\%\\
         & 20& 148.40\%& 141.01\%& 145.86\%& 151.93\%& 163.82\%\\\midrule
         \multirow{3}{*}{$\mathcal{O}^{t}_{self}$}& 5& 139.28\%& 140.44\%& 146.36\%& 144.84\%& 142.03\%\\ 
         & 10& 141.30\%& 143.09\%& 148.94\%& 148.39\%& 149.47\%\\
         & 20& 144.06\%& 146.45\%& 152.05\%& 153.11\%& 161.04\%\\
         \bottomrule\bottomrule
    \end{tabular}
    }
    \\\# Avg. Inter./U denotes average interactions per user.\\ \# Avg. Inter./I denotes average interactions per item. \\ \# Matrix Size calculated by \# Users times \# Items. \\ The calculation and approximation process of $\mathcal{O}_{avg}$ can be found in the Appendix~\ref{appendix:math}.
    \vskip -0.1in
\end{table}

For all $k$, the overlap rates $\mathcal{O}^{v}_{real}$ and $\mathcal{O}^{t}_{real}$ are significantly higher than the average overlap rate $\mathcal{O}_{avg}$. This indicates that items with modality information similar to those interacted with by users can effectively reflect user preferences. Notably, the calculation and approximation process of $\mathcal{O}_{avg}$ can be found in the Appendix~\ref{appendix:math}. Additionally, the overlap rate $\mathcal{O}^{vt}_{real}$ is higher than both $\mathcal{O}^{v}_{real}$ and $\mathcal{O}^{t}_{real}$, suggesting that items with similar modality information across multiple modalities are even more effective in capturing user preferences.

For all $k$, the overlap rates $\mathcal{O}^{v}_{self}$ and $\mathcal{O}^{t}_{self}$ are almost between 140\% and 160\%. Meanwhile, the interactions per user for all datasets are consistently around 8-10. This indicates that the items interacted with by users generally have high modality similarity, leading to a substantial amount of redundancy in the constructed similarity-aware virtual interactions. This suggests that users tend to have relatively fixed modality preferences.

From our investigation, we have the following observations:
\begin{itemize}[leftmargin=*]
    \item Items selected based on the modality similarity of each user's interacted items align well with the user's preferences.
    \item Multiple modal similarities have synergistic effects on inferring user preferences.
    \item Users always have relatively fixed modality preferences.
\end{itemize}

This investigation strongly supports our plug-and-play framework proposed in Section~\ref{sec:methodology}, which effectively builds virtual similarity-aware user-item interactions to mitigate the data sparsity problem.
\section{Methodology}
\label{sec:methodology}
\begin{figure*}
    \centering
    \includegraphics[width=1\linewidth]{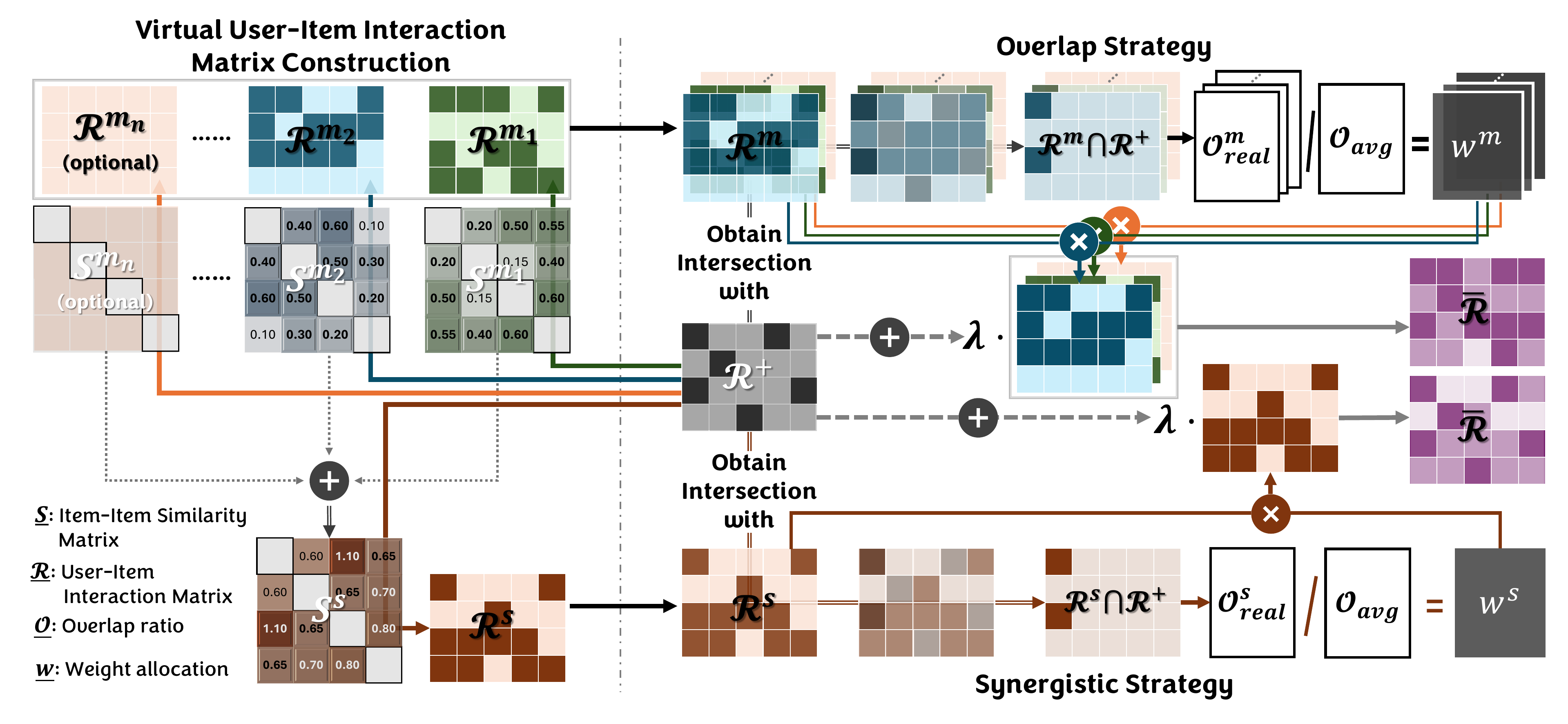}
     \vskip -0.15in
    \caption{The overall framework of our VI-MMRec.}
      \vskip -0.1in
    \label{fig:overview}
\end{figure*}

Motivated by the analyses in Section~\ref{sec:investigation}, we propose a plug-and-play framework, VI-MMRec, designed to enhance user-item interactions through similarity-aware virtual user-item interactions for multimodal recommendation. Notably, VI-MMRec is modality-agnostic, making it seamlessly compatible with any existing multimodal recommendation models. Moreover, VI-MMRec can be divided into two steps: 1) constructing similarity-aware virtual user-item interactions, and 2) augmenting user-item interactions via similarity-aware virtual user-item interactions. The overall framework of our proposed VI-MMRec is illustrated in Figure~\ref{fig:overview}.

\subsection{Constructing Virtual User-Item Interactions}
\label{subsec:build}
This part has been briefly introduced in Section~\ref{sec:investigation}. To maintain the logical coherence of our framework, we provide a formal and detailed description in this subsection. 

In Section~\ref{sec:investigation}, we analyzed that multiple modalities can work synergistically to infer user preferences. Based on this insight, we propose the following two strategies to construct similarity-aware virtual user-item interaction matrices:
\begin{itemize}[leftmargin=*]
    \item \textbf{Overlay Strategy:} For each modality, we independently construct a similarity-aware virtual user-item interaction matrix based on the similarity of modality features of items that users have interacted with in real user-item interactions. These matrices are subsequently combined through summation, where overlapping regions are assigned cumulative weights to enhance the augmentation of user-item interactions.
    \item \textbf{Synergistic Strategy:} We construct a single synergistic similarity-aware virtual user-item interaction matrix by considering the aggregated similarity of all modalities' features of items that users have interacted with in the real user-item interactions.
\end{itemize}

\subsubsection{Overlay Strategy} For each modality $m$, we compute a complete item similarity matrix using cosine similarity between all item pairs, formally defined as:
\begin{equation}
\label{eq:modality matrix}
    \mathcal{S}^m_{i,j}=\frac{(\mathbf{e}_{i^m})^{\top} \mathbf{e}_{j^m}}{\|\mathbf{e}_{i^m}\|\|\mathbf{e}_{j^m}\|},
\end{equation}
where $\mathbf{e}_{i^{m}}$ and $\mathbf{e}_{j^{m}}$ denote the row $i$ and row $j$ of item embedding $\mathbf{E}_{i^m}$ with modality $m$, respectively. For each interaction $\mathcal{R}_{u,i}$ in the real user-item interaction matrix $\mathcal{R}^+$, we identify a modality-specific item set $\mathcal{I}_{u,i}^m$ that retains the top-$k$ most similar items based on the similarity matrix $\mathcal{S}^m$. Specifically, for each $j \in \mathcal{I}_{u,i}^m$, the condition $\mathcal{S}^m_{i,j} \in \text { top-} k(\mathcal{S}^m_{i,p}|p \in \mathcal{I},p \neq i)$ must be satisfied. Using this, we construct the similarity-aware virtual user-item interaction matrix $\mathcal{R}^m$ for each modality $m$. Each similarity-aware virtual user-item interaction $\mathcal{R}^m_{u,j}$ satisfies $j \in \mathcal{I}^m_{u,i}$, where $\mathcal{R}_{u,i} \in \mathcal{R}^+$. This ensures that the virtual user-item interactions are derived based on the top-$k$ most similar items from the original user-item interactions, specific to the corresponding modality. This strategy enables the construction of virtual user-item interactions for each modality between users and items similar to those they have interacted with. Moreover, items that are similar across multiple modalities are assigned higher weights, reflecting their stronger relevance.

\subsubsection{Synergistic Strategy:} We compute a synergistic complete item similarity matrix using cosine similarity between all item pairs, formally defined as:
\begin{equation}
\label{eq:synergistic matrix}
\mathcal{S}_{i,j}^{s}=\sum_{m}^{\mathcal{M}}\frac{(\mathbf{e}_{i^m})^{\top} \mathbf{e}_{j^m}}{\|\mathbf{e}_{i^m}\|\|\mathbf{e}_{j^m}\|}.
\end{equation}

For each interaction $\mathcal{R}_{u,i}$ in the real user-item interaction matrix $\mathcal{R}^+$, we identify a synergistic item set $\mathcal{I}_{u,i}$ that retains the top-$k$ most similar items based on the similarity matrix $\mathcal{S}^{s}$. Specifically, for each $j \in \mathcal{I}_{u,i}$, the condition $\mathcal{S}_{i,j}^{s} \in \text { top-} k(\mathcal{S}_{i,p}^{s}|p \in \mathcal{I},p \neq i)$ must be satisfied. Using this, we construct a synergistic virtual user-item interaction matrix\footnote{$\mathcal{R}^{vt}$ in Section~\ref{sec:investigation} is an instance of $\mathcal{R}^{s}$, which contains visual and textual modalities.} $\mathcal{R}^{s}$. Each synergistic virtual interaction $\mathcal{R}_{u,j}$ satisfies $j \in \mathcal{I}_{u,i}$, where $\mathcal{R}_{u,i} \in \mathcal{R}^+$. This strategy ensures that the virtual user-item interactions are derived by considering top-$k$ most similar items across all modalities, integrating their combined effects into the interaction matrix. However, this strategy inevitably ignores some strongly related items in a single modality that match user preferences. We discuss the effectiveness of these two strategies empirically in Section~\ref{subsec:RQ1-2}.

\subsection{Augmenting User-item Interactions}
After constructing similarity-aware virtual user-item interactions, we will introduce how to augment the real user-item interaction matrix with our virtual user-item interaction matrices. Treating virtual user-item interactions and real user-item interactions with equal weight is arbitrary, and even assigning equal weight to virtual user-item interactions from different modalities is unreasonable. To address this, we propose a statistically informed weight allocation strategy, enabling us to reasonably assign appropriate weights to virtual user-item interactions from different modalities.

For overlay strategy, as discussed in Section~\ref{sec:investigation}, a higher overlap rate $\mathcal{O}^{m}_{real}$ between the similarity-aware virtual user-item interaction matrix $\mathcal{R}^{m}$ and the real user-item interaction matrix $\mathcal{R}^{+}$ indicates that the items interacted with by users exhibit high similarity in modality $m$. This implies that the features of modality $m$ can effectively capture and reflect user preferences. However, as shown in Table~\ref{tab:investigation}, the data sparsity varies across different datasets, making it insufficient to directly determine the weight of modality $m$ solely based on $\mathcal{O}^{m}_{real}$, as it cannot fully represent the correlation between modality $m$ and user preferences. Therefore, we further propose a statistically informed weight allocation mechanism to refine the weight allocation for each modality by measuring the ratio of $\mathcal{O}^{m}_{real}$ to the overlap $\mathcal{O}_{avg}$, where $\mathcal{O}_{avg}$ is the overlap between the real user-item interaction matrix $\mathcal{R}^{+}$ and virtual user-item interactions constructed randomly without considering modality similarity. Formally, 
\begin{equation}
\label{eq:weightm}
    \mathbf{w}^{m} = \frac{\mathcal{O}^{m}_{real}}{\mathcal{O}_{avg}},
\end{equation}
where ratio $\mathbf{w}^{m}$ provides a more accurate representation of the relevance of modality $m$ to user preferences. The final augmented user-item interaction matrix $\bar{\mathcal{R}}$ can be calculated as:
\begin{equation}
\label{eq:matrixm}
    \bar{\mathcal{R}} = \operatorname{Confine}(\mathcal{R} + \lambda\sum_{m}^{\mathcal{M}}\mathbf{w}^{m}\mathcal{R}^{m}),
\end{equation}
where $\operatorname{Confine}(\cdot)$ ensures that the weight of each interaction is confined within the range $[0,1]$. Formally, for any interaction where $(\mathcal{R}_{i,j} + \lambda\sum_{m}^{\mathcal{M}}\mathbf{w}^{m}\mathcal{R}_{i,j}^{m}) > 1$, we set $\bar{\mathcal{R}}_{i,j} = 1$. $\lambda$ is a hyper-parameter to control the weight of virtual user-item interactions.

For synergistic strategy, we get refined weight allocation for synergistic features by measuring the ratio of $\mathcal{O}^{s}_{real}$ to the overlap $\mathcal{O}_{avg}$, where $\mathcal{O}_{avg}$ is the overlap between the real user-item interaction matrix $\mathcal{R}$ and virtual user-item interactions constructed randomly without considering modality similarity. Formally, 
\begin{equation}
\label{eq:weights}
    \mathbf{w}^{s} = \frac{\mathcal{O}^{s}_{real}}{\mathcal{O}_{avg}},
\end{equation}
where ratio $\mathbf{w}^{m}$ provides a more accurate representation of the relevance of multimodal synergistic features to user preferences. The final augmented user-item interaction matrix $\bar{\mathcal{R}}$ can be calculated as:
\begin{equation}
\label{eq:matrixs}
    \bar{\mathcal{R}} = \operatorname{Confine}(\mathcal{R} + \lambda\mathbf{w}^{s}\mathcal{R}^{s}),
\end{equation}
where $\operatorname{Confine}(\cdot)$ ensures that the weight of each interaction is confined within the range $[0,1]$. $\lambda$ is a hyperparameter to control the weight of virtual user-item interactions. In Section~\ref{subsec:RQ4}, we provide an empirical analysis for the impact of different hyper-parameter $\lambda$.

\subsection{Training Cost-free Analysis}
We conduct an in-depth analysis of the remarkable cost-free advantage of our VI-MMRec framework during the training phase and further illustrate how it can be seamlessly compatible with any existing multimodal recommendation models. 

Specifically, the modality embedding $\mathbf{e}_{i}^{m}$ for each item $i$ is extracted using a pre-trained model and remains frozen during the training process of the recommendation model \cite{zhou2023tale,xu2025survey,zhou2023mmrecsm}. This allows us to construct the augmented virtual user-item interaction matrix $\bar{\mathcal{R}}$ prior to training. For any given dataset, $\bar{\mathcal{R}}$ only needs to be generated once and can be stored for reuse across different models. Furthermore, since the augmented interaction matrix $\bar{\mathcal{R}} \in \mathbb{R}^{|\mathcal{U}| \times |\mathcal{I}|}$ has the same dimensionality as original user-item interaction matrix $\mathcal{R}$, it introduces no additional computational overhead. Therefore, our VI-MMRec has a remarkable cost-free advantage during the training phase. For any existing multimodal recommendation model, we can achieve satisfactory performance improvements by simply replacing its original $\mathcal{R}$ with $\bar{\mathcal{R}}$. This has been empirically validated in Section~\ref{sec:experiments}. We present the overall procedure in Algorithm~\ref{alg:framework}.

\begin{algorithm}
\caption{Procedure for VI-MMRec}
\label{alg:framework}
\KwIn{Item embedding $\mathbf{E}_{i}^{m}$ with modality $m$, which extracted by pre-trained encoders. Original user-item interaction matrix $\mathcal{R}$.}
\If{Overlap Strategy}{
    $\#$ Overlap Strategy\\
    Construct item similarity matrix $\mathcal{S}^{m}$ for each modality $m$ via Eq.~\ref{eq:modality matrix}.\\
    Get similarity-aware virtual user-item interaction matrix $\mathcal{R}^m$ for each modality $m$ via $\mathcal{S}^{m}$.\\
    Calculate weight $\mathbf{w}^{m}$ for each modality $m$ via Eq.~\ref{eq:modality overlap} and Eq.~\ref{eq:weightm}.
    Get the final augmented user-item interaction matrix $\bar{\mathcal{R}}$ via Eq.~\ref{eq:matrixm}.
}
\Else{ 
     $\#$ Synergistic Strategy\\
    Construct synergistic item similarity matrix $\mathcal{S}^{s}$ via Eq.~\ref{eq:synergistic matrix}.\\
    Get similarity-aware virtual user-item interaction matrix $\mathcal{R}^s$ for synergistic modalities via $\mathcal{S}^{s}$.\\
    Calculate weight $\mathbf{w}^{s}$ for synergistic modalities via Eq.~\ref{eq:synergistic overlap} and Eq.~\ref{eq:weights}.
    Get the final augmented user-item interaction matrix $\bar{\mathcal{R}}$ via Eq.~\ref{eq:matrixs}.
}
Using $\bar{\mathcal{R}}$ to replace $\mathcal{R}$ for each datasets.\\
Model training (Model-agnostic).

\end{algorithm}
 \vskip -0.1in
\section{Experiments}
\label{sec:experiments}

\begin{table*}[!ht]
    \centering
\caption{Statistics of all experimented datasets with multimodal item contents.}
 \vskip -0.15in
\label{tab:dataset_statistics}
    \begin{tabular}{cccccccccccccccccccc}
    \hline
         Dataset&& \multicolumn{2}{c}{Baby} && \multicolumn{2}{c}{Sports} && \multicolumn{2}{c}{Clothing} && \multicolumn{2}{c}{Pet} && \multicolumn{2}{c}{Office} && \multicolumn{3}{c}{Tiktok}\\
         \hline \hline
         Modality && V & T && V & T && V & T && V & T && V & T && V & T & A\\ 
         Dim. && 4096 & 384 && 4096 & 384 && 4096 & 384 && 4096 & 384 && 4096 & 384 && 128 & 768 & 128 \\ \cline{1-1} \cline{3-4} \cline{6-7} \cline{9-10} \cline{12-13} \cline{15-16} \cline{18-20}
         User && \multicolumn{2}{c}{19,445} && \multicolumn{2}{c}{35,598} && \multicolumn{2}{c}{39,387} && \multicolumn{2}{c}{19,856} && \multicolumn{2}{c}{4,905} && \multicolumn{3}{c}{9,319} \\
         Item && \multicolumn{2}{c}{7,050} && \multicolumn{2}{c}{18,357} && \multicolumn{2}{c}{23,033} && \multicolumn{2}{c}{8,510} && \multicolumn{2}{c}{2,420} && \multicolumn{3}{c}{6,710} \\
         Interaction && \multicolumn{2}{c}{160,792} && \multicolumn{2}{c}{296,337} && \multicolumn{2}{c}{278,677} && \multicolumn{2}{c}{157,836} && \multicolumn{2}{c}{53,258} && \multicolumn{3}{c}{59,541} \\ \hline
         Sparsity && \multicolumn{2}{c}{99.88\%} && \multicolumn{2}{c}{99.95\%} && \multicolumn{2}{c}{99.97\%} && \multicolumn{2}{c}{99.91\%} && \multicolumn{2}{c}{99.55\%} && \multicolumn{3}{c}{99.90\%} \\
         \hline
    \end{tabular}
     \vskip -0.1in
\end{table*}
\begin{table*}[!ht]
\caption{Performance comparison of baselines with or without VI-MMRec on all datasets in terms of Recall@10 and NDCG@10. The superscript $^*$ indicates the improvement is statistically significant where the p-value is less than 0.01.}
 \vskip -0.15in
\centering
\tabcolsep=0.04in
\label{tab:comparison results}
\resizebox{\linewidth}{!}{
    \begin{tabular}{|c|cc|cc|cc|cc|cc|cc|}
    \hline
         Datasets&  \multicolumn{2}{c|}{Baby}&  \multicolumn{2}{c|}{Sports}&  \multicolumn{2}{c|}{Clothing}&  \multicolumn{2}{c|}{Pet}&  \multicolumn{2}{c|}{Office}& \multicolumn{2}{c|}{TikTok}\\\hline
         Metrics& Recall@10& NDCG@10& Recall@10& NDCG@10& Recall@10& NDCG@10& Recall@10& NDCG@10& Recall@10& NDCG@10& Recall@10& NDCG@10\\\hline
         MMGCN & 0.0378& 0.0200& 0.0370& 0.0193& 0.0218& 0.0110& 0.0619& 0.0329& 0.0560& 0.0305& 0.0463& 0.0231\\
         +VI-MMRec-O& \textbf{0.0412$^*$}& \textbf{0.0216$^*$}& \textbf{0.0414$^*$}& \textbf{0.0215$^*$}& \textbf{0.0236$^*$}& \textbf{0.0119$^*$}& \textbf{0.0712$^*$}& \textbf{0.0376$^*$}& \textbf{0.0633$^*$}& \textbf{0.0348$^*$}& \textbf{0.0566$^*$}& \textbf{0.0278$^*$}\\
         +VI-MMRec-S& 0.0401& 0.0211& 0.0400& 0.0208& 0.0231& 0.0117& 0.0689& 0.0366& 0.0615& 0.0336& 0.0539& 0.0268\\
         \gray Improv. &\gray 8.99\% &\gray 8.00\% &\gray 11.89\% &\gray 11.40\% &\gray 8.26\% &\gray 8.18\% &\gray 15.02\% &\gray 14.29\% &\gray 13.04\% &\gray 14.10\% &\gray 22.25\% &\gray 20.35\%\\\hline
         DualGNN & 0.0448& 0.0240& 0.0568& 0.0310& 0.0454& 0.0241& 0.0902& 0.0503& 0.0873& 0.0477& 0.0552& 0.0278\\
         +VI-MMRec-O& \textbf{0.0485$^*$}& \textbf{0.0262$^*$}& \textbf{0.0609$^*$}& \textbf{0.0335$^*$}& \textbf{0.0487$^*$}& \textbf{0.0258$^*$}& \textbf{0.0985$^*$}& \textbf{0.0548$^*$}& \textbf{0.0950$^*$}& \textbf{0.0512$^*$}& \textbf{0.0636$^*$}& \textbf{0.0318$^*$}\\
         +VI-MMRec-S& 0.0473& 0.0256& 0.0595& 0.0326& 0.0479& 0.0253& 0.0963& 0.0534& 0.0928& 0.0498& 0.0609& 0.0305\\
         \gray Improv. &\gray 8.26\% &\gray 9.17\% &\gray 7.22\% &\gray 8.06\% &\gray 7.27\% &\gray 7.05\% &\gray 9.20\% &\gray 8.95\% &\gray 8.82\% &\gray 7.34\% &\gray 15.22\% &\gray 14.39\%\\\hline
         SLMRec & 0.0529& 0.0290& 0.0663& 0.0365& 0.0452& 0.0247& 0.0976& 0.0533& 0.0761& 0.0414& 0.0503& 0.0251\\
         +VI-MMRec-O& \textbf{0.0565$^*$}& \textbf{0.0308$^*$}& \textbf{0.0710$^*$}& \textbf{0.0390$^*$}& \textbf{0.0489$^*$}& \textbf{0.0265$^*$}& \textbf{0.1038$^*$}& \textbf{0.0564$^*$}& \textbf{0.0828$^*$}& \textbf{0.0454$^*$}& \textbf{0.0571$^*$}& \textbf{0.0287$^*$}\\
         +VI-MMRec-S& 0.0554& 0.0303& 0.0694& 0.0378& 0.0482& 0.0261& 0.1022& 0.0556& 0.0809& 0.0441& 0.0548& 0.0277\\
         \gray Improv. &\gray 6.81\% &\gray 6.21\% &\gray 7.09\% &\gray 6.85\% &\gray 8.19\% &\gray 7.29\% &\gray 6.35\% &\gray 5.82\% &\gray 8.80\% &\gray 9.66\% &\gray 13.52\% &\gray 14.34\%\\\hline
         FREEDOM & 0.0627& 0.0330& 0.0717& 0.0385& 0.0629& 0.0341& 0.1086& 0.0595& 0.0952& 0.0518& 0.0589& 0.0295\\
         +VI-MMRec-O& \textbf{0.0663$^*$}& \textbf{0.0351$^*$}& \textbf{0.0755$^*$}& \textbf{0.0407$^*$}& \textbf{0.0662$^*$}& \textbf{0.0365$^*$}& \textbf{0.1143$^*$}& \textbf{0.0626$^*$}& \textbf{0.1006$^*$}& \textbf{0.0547$^*$}& \textbf{0.0663$^*$}& \textbf{0.0330$^*$}\\
         +VI-MMRec-S& 0.0653& 0.0343& 0.0743& 0.0399& 0.0652& 0.0357& 0.1127& 0.0616& 0.0992& 0.0540& 0.0639& 0.0318\\
         \gray Improv. &\gray 5.74\% &\gray 6.36\% &\gray 5.30\% &\gray 5.71\% &\gray 5.25\% &\gray 7.04\% &\gray 5.25\% &\gray 5.21\% &\gray 5.67\% &\gray 5.60\% &\gray 12.56\% &\gray 11.86\%\\\hline
         DRAGON & 0.0662& 0.0345& 0.0752& 0.0413& 0.0671& 0.0365& 0.1151& 0.0630& 0.1014& 0.0557& 0.0682& 0.0341\\
         +VI-MMRec-O& \textbf{0.0699$^*$}& \textbf{0.0368$^*$}& \textbf{0.0790$^*$}& \textbf{0.0437$^*$}& \textbf{0.0714$^*$}& \textbf{0.0388$^*$}& \textbf{0.1210$^*$}& \textbf{0.0664$^*$}& \textbf{0.1076$^*$}& \textbf{0.0594$^*$}& \textbf{0.0750$^*$}& \textbf{0.0378$^*$}\\
         +VI-MMRec-S& 0.0689& 0.0359& 0.0779& 0.0430& 0.0702& 0.0382& 0.1193& 0.0652& 0.1059& 0.0582& 0.0722& 0.0367\\
         \gray Improv. &\gray 5.59\% &\gray 6.67\% &\gray 5.05\% &\gray 5.81\% &\gray 6.41\% &\gray 6.30\% &\gray 5.13\% &\gray 5.40\% &\gray 6.11\% &\gray 6.64\% &\gray 9.97\% &\gray 10.85\%\\\hline
         LGMRec & 0.0647& 0.0333& 0.0719& 0.0387& 0.0555& 0.0302& 0.1057& 0.0584& 0.0959& 0.0514& 0.0610& 0.0304\\
         +VI-MMRec-O& \textbf{0.0681$^*$}& \textbf{0.0353$^*$}& \textbf{0.0760$^*$}& \textbf{0.0412$^*$}& \textbf{0.0592$^*$}& \textbf{0.0321$^*$}& \textbf{0.1123$^*$}& \textbf{0.0619$^*$}& \textbf{0.1018$^*$}& \textbf{0.0547$^*$}& \textbf{0.0674$^*$}& \textbf{0.0336$^*$}\\
         +VI-MMRec-S& 0.0671& 0.0348& 0.0747& 0.0404& 0.0581& 0.0313& 0.1104& 0.0610& 0.1003& 0.0540& 0.0651& 0.0326\\
         \gray Improv. &\gray 5.26\% &\gray 6.01\% &\gray 5.70\% &\gray 6.46\% &\gray 6.67\% &\gray 6.29\% &\gray 6.24\% &\gray 5.99\% &\gray 6.15\% &\gray 6.42\% &\gray 10.49\% &\gray 10.53\%\\\hline
         DiffMM& 0.0643& 0.0332& 0.0731& 0.0395& 0.0556& 0.0310& 0.1060& 0.0582& 0.0975& 0.0525& 0.0672& 0.0335\\
         +VI-MMRec-O& \textbf{0.0678$^*$}& \textbf{0.0351$^*$}& \textbf{0.0771$^*$}& \textbf{0.0417$^*$}& \textbf{0.0588$^*$}& \textbf{0.0329$^*$}& \textbf{0.1121$^*$}& \textbf{0.0616$^*$}& \textbf{0.1026$^*$}& \textbf{0.0553$^*$}& \textbf{0.0748$^*$}& \textbf{0.0373$^*$}\\
         +VI-MMRec-S& 0.0667& 0.0343& 0.0759& 0.0411& 0.0577& 0.0323& 0.1105& 0.0607& 0.1010& 0.0544& 0.0727& 0.0364\\
         \gray Improv. &\gray 5.44\% &\gray 5.72\% &\gray 5.47\% &\gray 5.57\% &\gray 5.76\% &\gray 6.13\% &\gray 5.75\% &\gray 5.84\% &\gray 5.23\% &\gray 5.33\% &\gray 11.31\% &\gray 11.34\%\\\hline
    \end{tabular}
    }
     \vskip -0.1in
\end{table*}

We conduct extensive experiments on our VI-MMRec, aiming to answer the following research questions (RQs): \textbf{RQ1:} Can VI-MMRec enhance the performance of multimodal recommendations? \textbf{RQ2:} How effective are the different strategies in VI-MMRec? \textbf{RQ3:} How do the key components in VI-MMRec affect performance enhancement? \textbf{RQ4:} Can VI-MMRec enhance multimodal recommendations in different sparse data scenarios? \textbf{RQ5:} Can VI-MMRec enhance multimodal recommendations in cold-start settings? \textbf{RQ6:} How does VI-MMRec perform under representation noise and information error? \textbf{RQ7:} What is the impact of key hyper-parameters in VI-MMRec?

\subsection{Experimental Settings}
\subsubsection{Datasets}
The experiments are conducted on six real-world datasets, five containing two modalities: Baby, Sports, Clothing, Pet, and Office from the Amazon dataset \cite{mcauley2015image}. These datasets include textual and visual features, derived from item descriptions and corresponding images. The data preprocessing for these datasets follows the methodology outlined in MMRec \cite{zhou2023mmrecsm}. To further evaluate the performance of VI-MMRec in scenarios involving multiple modalities, we also conduct experiments on the TikTok dataset \cite{jiang2024diffmm}, which consists of user interaction logs with short videos collected from the TikTok platform. For the TikTok dataset, our data processing follows previous works \cite{jiang2024diffmm,wei2023multi}. Table~\ref{tab:dataset_statistics} shows the statistics of these datasets.

\subsubsection{Evaluation Protocols} 
To evaluate the performance fairly, we adopt two widely used metrics: Recall@10 and NDCG@10. We report the average metrics of all users in the test dataset. We follow the popular evaluation setting \cite{xu2024mentor,zhou2023tale,guo2024lgmrec} with a random data splitting 8:1:1 for training, validation, and testing.

\subsubsection{Baselines}
We examine the performance of our VI-MMRec across 7 advanced multimodal recommendation models: \textbf{MMGCN} \cite{wei2019mmgcn}, \textbf{DualGNN} \cite{wang2021dualgnn}, \textbf{SLMRec} \cite{tao2022self}, \textbf{FREEDOM} \cite{zhou2023tale}, \textbf{DRAGON} \cite{zhou2023enhancing}, \textbf{LGMRec} \cite{guo2024lgmrec}, and \textbf{DiffMM} \cite{jiang2024diffmm}. Details can be found in the Appendix~\ref{appendix:baseline}.

\subsubsection{Implementation Details}
We retain the standard settings for all baselines and fix batch size $B = 2048$. For VI-MMRec, we apply a grid search on hyper-parameters $\lambda$ in $\{1e^{-3}, 5e^{-3}, 1e^{-2}, 5e^{-2}\}$, and the value $k$ for constructing virtual user-item interactions in $\{5, 10, 20\}$. All models are implemented in PyTorch, using the Adam optimizer \cite{kingma2014adam} and Xavier initialization \cite{glorot2010understanding} with default parameters, and all training and evaluation are conducted on a single RTX4090 GPU. Moreover, we implement two instances of VI-MMRec with different strategies for constructing virtual user-item interaction matrices: a) VI-MMRec-O, which adopts an overlay strategy for our VI-MMRec. b) VI-MMRec-S, which adopts a synergistic strategy for our VI-MMRec.

\subsection{Overall Performance (RQ1 \& RQ2)}
\label{subsec:RQ1-2}
We evaluate the effectiveness of our VI-MMRec on various models for multimodal recommendation scenarios and to discern the impact of strategies for constructing virtual user-item interaction matrices. From Table~\ref{tab:comparison results}, we find the following observations:

\noindent \textbf{Observation1:}
Both VI-MMRec-O and VI-MMRec-S effectively enhance the performance of various multimodal recommendation models. As shown in Table \ref{tab:comparison results}, we conducted extensive experiments adopting VI-MMRec-O and VI-MMRec-S on seven multimodal recommendation models across six different public datasets. The results demonstrate that both VI-MMRec-O and VI-MMRec-S achieve significant improvements over all baselines across all evaluation metrics. In summary, the experimental findings validate that constructing a virtual user-item interaction matrix based on the similarity of modality features from items, which users have interacted with in real interactions, effectively alleviates the data sparsity problem and improves recommendation performance. 

\noindent \textbf{Observation2:}
All models with VI-MMRec-O and VI-MMRec-S achieve higher performance improvement on the TikTok dataset than on other datasets. We attribute this phenomenon to the TikTok dataset containing three modalities, as the presence of more modalities allows the constructed virtual user-item interaction matrix to more accurately infer user preferences.

\noindent \textbf{Observation3:}
As shown in Table \ref{tab:comparison results}, VI-MMRec-O consistently outperforms VI-MMRec-S on all models across all datasets. We attribute this phenomenon to the synergistic strategy (VI-MMRec-S) focusing solely on items with high similarity across all modalities while neglecting items where a single modality plays a decisive role. This limits the constructed virtual user-item interaction matrix to infer user preferences accurately.

\begin{figure}[!h]
    \centering
    \vskip -0.1in
    \includegraphics[width=1\linewidth]{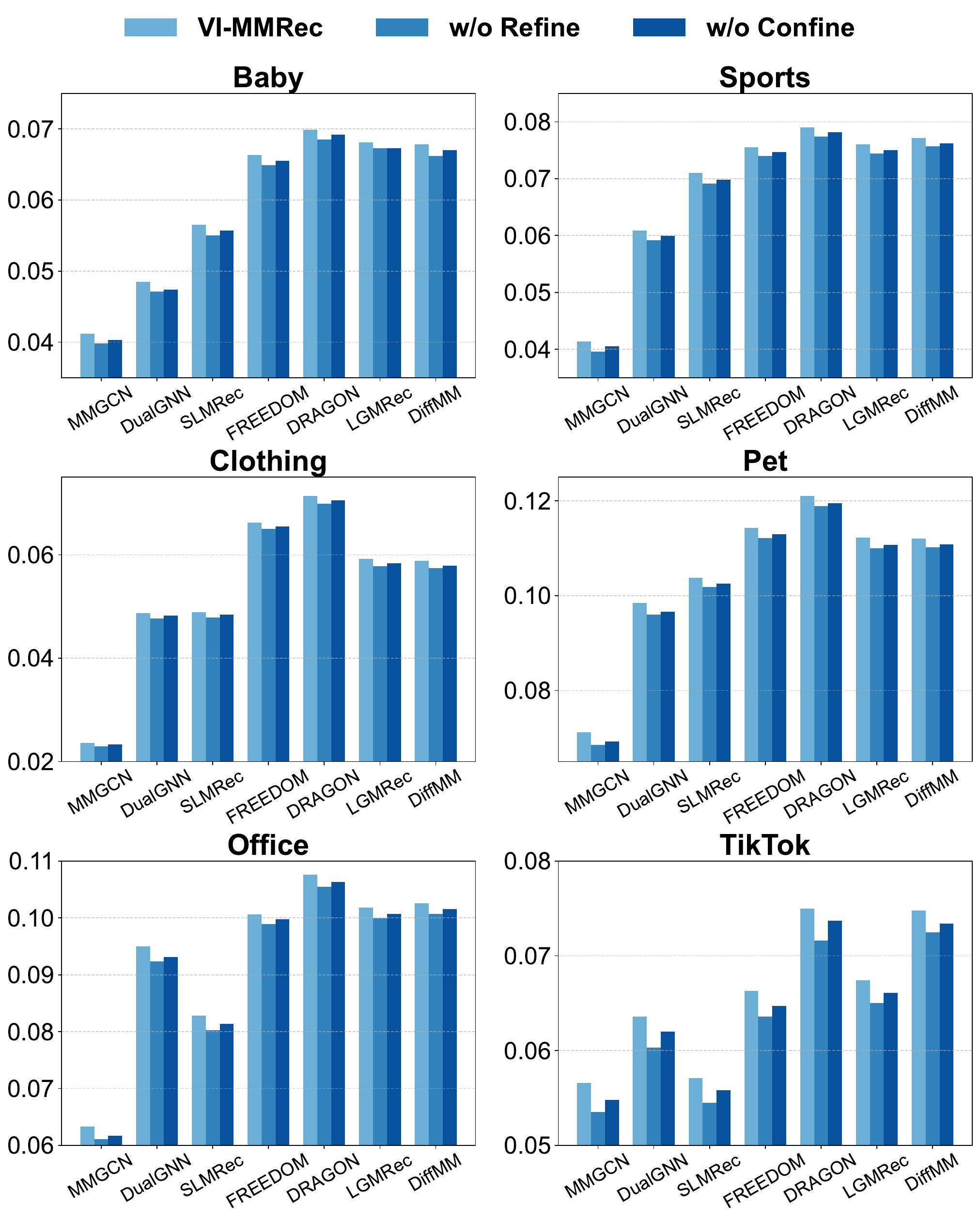}
     \vskip -0.15in
    \caption{Performance comparison for VI-MMRec and all variants on six multimodal recommendation models across all six datasets regarding Recall@10.}
    \label{fig:ablation}
     \vskip -0.1in
\end{figure}

\subsection{Ablation Study (RQ3)}
\label{subsec:RQ3}
To validate the effectiveness of VI-MMRec, we conduct experiments to justify the importance of key components. We design the following variants:
a) \textbf{$w/o$ Refine} removes refinement of weight allocation in Eq.~\ref{eq:weightm} and Eq.~\ref{eq:weights}. b) \textbf{$w/o$ Confine} removes confinement of edge weight in Eq.~\ref{eq:matrixm} and Eq.~\ref{eq:matrixs}. The results are demonstrated in Figure~\ref{fig:ablation}. All the components contribute to the performance of VI-MMRec. Specifically, we have the following observations: 1) VI-MMRec achieves higher performance than \textbf{$w/o$ Refine}, demonstrating the necessity of refining the weight allocation based on dataset statistics. 2) The performance of \textbf{$w/o$ Confine} is significantly lower than that of VI-MMRec, which highlights the importance of constraining edge weights within a suitable and realistic range.

\begin{figure*}[!h]
    \centering
    \vskip -0.1in
    \includegraphics[width=0.95\linewidth]{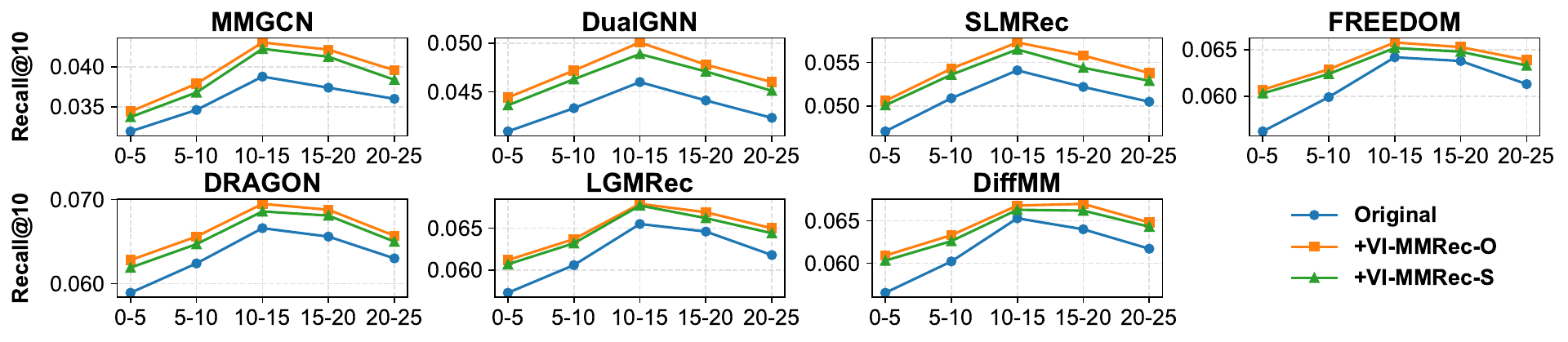}
    \vskip -0.15in
    \caption{Sparsity degree analysis on the Baby dataset in terms of Recall@10.}
    \vskip -0.1in
    \label{fig:sparsity}
\end{figure*}

\begin{table}[!ht]
    \centering
    \caption{Cold start analysis across Baby and Tiktok datasets.}
    \label{tab:cold-start}
     \vskip -0.15in
    \resizebox{\linewidth}{!}{
    \begin{tabular}{|c|cc|cc|}
    \hline
        Datasets & \multicolumn{2}{c|}{Baby} & \multicolumn{2}{c|}{TikTok} \\ \hline
        Metrics & Recall@10 & NDCG@10 & Recall@10 & NDCG@10 \\ \hline
        MMGCN & 0.0103 & 0.0062 & 0.0214 & 0.0112 \\ 
        +VI-MMRec-O & \textbf{0.0120} & \textbf{0.0071} & \textbf{0.0247} & \textbf{0.0130} \\ 
        +VI-MMRec-S & 0.0117 & 0.0069 & 0.0244 & 0.0127 \\ \hline
        DualGNN & 0.0132 & 0.0077 & 0.0241 & 0.0127 \\ 
        +VI-MMRec-O & \textbf{0.0155} & \textbf{0.0090} &\textbf{0.0276} & \textbf{0.0143} \\ 
        +VI-MMRec-S & 0.0152 & 0.0088 & 0.0271 & 0.0139 \\ \hline
        SLMRec & 0.0172 & 0.0101 & 0.0230 & 0.0117 \\
        +VI-MMRec-O & \textbf{0.0203} & \textbf{0.0120} & \textbf{0.0253} & \textbf{0.0129} \\ 
        +VI-MMRec-S & 0.0198 & 0.0117 & 0.0249 & 0.0126 \\ \hline
        FREEDOM & 0.0348 & 0.0195 & 0.0301 & 0.0149 \\ 
        +VI-MMRec-O & \textbf{0.0385} & \textbf{0.0216} & \textbf{0.0330} & \textbf{0.0162} \\ 
        +VI-MMRec-S & 0.0377 & 0.0211 & 0.0322 & 0.0157 \\ \hline
        DRAGON & 0.0378 & 0.0212 & 0.0402 & 0.0205 \\
        +VI-MMRec-O & \textbf{0.0412} & \textbf{0.0233} & \textbf{0.0437} & \textbf{0.0219} \\ 
        +VI-MMRec-S & 0.0404 & 0.0228 & 0.0429 & 0.0213 \\ \hline
        LGMRec & 0.0371 & 0.0208 & 0.0333 & 0.0169 \\ 
        +VI-MMRec-O & \textbf{0.0404} & \textbf{0.0226} & \textbf{0.0362} & \textbf{0.0182} \\ 
        +VI-MMRec-S & 0.0397 & 0.0220 & 0.0351 & 0.0175 \\ \hline
        DiffMM & 0.0336 & 0.0193 & 0.0345 & 0.0170 \\ 
        +VI-MMRec-O & \textbf{0.0367} & \textbf{0.0210} & \textbf{0.0380} & \textbf{0.0190} \\ 
        +VI-MMRec-S & 0.0360 & 0.0204 & 0.0372 & 0.0185 \\ \hline
    \end{tabular}
    }
\end{table}

\begin{figure*}[!h]
\vskip -0.05in
    \centering
    \subfigure[Hyper-parameter $k$] {
        \label{fig:recall10_k}
        \includegraphics[width=0.95\linewidth]{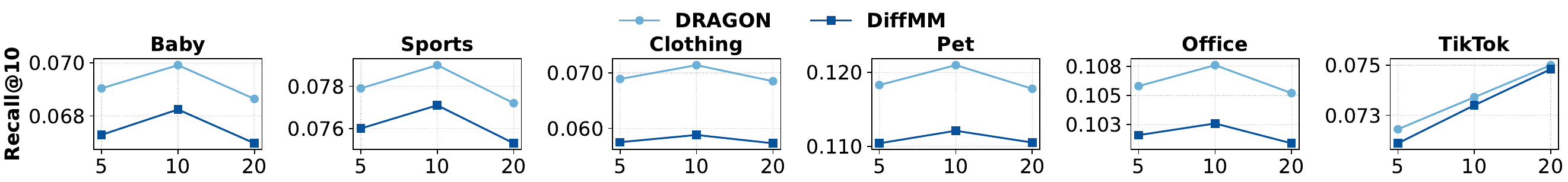}
        }
     \vskip -0.1in
    \subfigure[Hyper-parameter $\lambda$] {
        \label{fig:recall10_lambda}
        \includegraphics[width=0.95\linewidth]{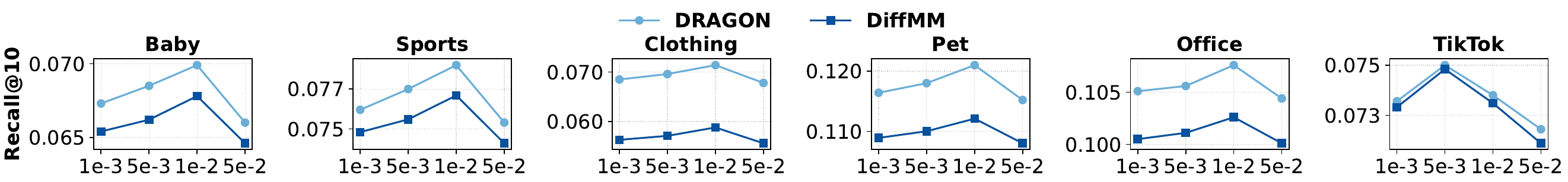}
        }
    \vskip -0.2in
    \caption{Performance comparison $w.r.t.$ key hyper-parameters ($k$ and $\lambda$) for DRAGON and DiffMM models across all datasets.}
    \label{fig:hyper}
    \vskip -0.1in
\end{figure*}

\subsection{Data Sparsity Analysis (RQ4)}
To evaluate the effectiveness of integrating VI-MMRec into advanced multimodal recommendation models under various data sparsity scenarios, we conduct experiments on subsets of the Baby dataset with different sparsity levels. To analyze the impact of data sparsity, we divide users into groups based on their interaction counts in the training set (e.g., the first group includes users who interacted with 1-5 items). As shown in Figure~\ref{fig:sparsity}, VI-MMRec consistently improves the performance of these models across all sparsity levels, validating its effectiveness in diverse sparse scenarios.

\subsection{Cold Start Analysis (RQ5)}
We present results on the cold-start scenario across Baby and TikTok datasets (following widely used settings \cite{zhang2022latent}). The experimental results in Table~\ref{tab:cold-start} show that both variants of VI-MMRec can improve the performance of multimodal recommendation models in cold-start scenarios. We attribute this to our virtual user-item interactions, which directly and explicitly reduce data sparsity, thereby enhancing the alignment between multimodal information and user behavior.

\subsection{Noise and Error Settings (RQ6)}
We considered both representation noise and information error. The former adds noise to the representation and the latter directly modifies the original multimodal data.

For representation noise, we directly inject noise into the raw modality representation. We build the following levels: L1 injected Gaussian noise  $\epsilon \sim \mathcal{N}(0,10^{-6})$, L2 injected Gaussian noise  $\epsilon \sim \mathcal{N}(0,10^{-5})$, L3 injected Gaussian noise  $\epsilon \sim \mathcal{N}(0,10^{-4})$, 

For information error, we random exchange two item's multimodal data. We build the following levels: L1 randomly exchange occurs with a probability of 1\%, L2 randomly exchange occurs with a probability of 3\%, and L3 randomly exchange occurs with a probability of 5\%. 

Table~\ref{tab:noise_error} shows experimental results on the Baby dataset, demonstrating that the performance gain of VI-MMRec remains stable and even enhances the model's robustness to noise. This is because the construction of virtual interactions is based on item similarity. Identifying modality-similar items in the dataset is not a strict 0/1 problem but rather involves a range of similar items. Therefore, a moderate amount of noise does not hinder the ability of virtual interactions to assist in learning user preferences.
\begin{table*}[h]
    \centering
    \caption{The terms "N" and "E" refer to representation noise and information error, respectively. Experiments conducted on the Baby dataset, in terms of Recall@10. "$\downarrow$" denotes the relative decrease compared to the original result after simulating risk.}
    \vskip -0.05in
\label{tab:noise_error}
\resizebox{1\linewidth}{!}{
\begin{tabular}{|c|c|cc|cc|cc||c|cc|cc|cc|}\hline
    Model& Ori.& N (L1)& $\downarrow$& N (L2)& $\downarrow$& N (L3)& $\downarrow$ & Ori.& E (L1)& $\downarrow$& E (L2)& $\downarrow$& E (L3)& $\downarrow$\\\hline
    \multirow{1}{*}{MMGCN}& 0.0378 & 0.0355 & 6.08\% & 0.0348 & 7.94\% & 0.0334 & 11.64\%& 0.0378 & 0.0349 & 7.67\% & 0.0341 & 9.79\% & 0.0329 & 13.00\% \\
    +VI-MMRec-O& 0.0412 & 0.0391 & 5.10\% & 0.0385 & 6.55\% & 0.0372 & 9.71\%  &0.0412 & 0.0387 & 6.07\% & 0.0383 & 7.04\% & 0.0371 & 9.95\% \\
    +VI-MMRec-S& 0.0401 & 0.0380 & 5.24\% & 0.0375 & 6.48\% & 0.0362 & 9.73\%  &0.0401 & 0.0376 & 6.23\% & 0.0373 & 6.98\% & 0.0361 & 9.98\%\\ \hline
    \multirow{1}{*}{FREEDOM}& 0.0627 & 0.0596 & 4.94\% & 0.0591 & 5.74\% & 0.0584 & 6.86\%  & 0.0627 & 0.0596 & 4.94\% & 0.0591 & 5.74\% & 0.0584 & 6.86\% \\
    +VI-MMRec-O& 0.0663 & 0.0637 & 3.92\% & 0.0630 & 4.98\% & 0.0623 & 6.03\%  & 0.0663 & 0.0637 & 3.92\% & 0.0630 & 4.98\% & 0.0624 & 5.88\% \\
    +VI-MMRec-S& 0.0653 & 0.0626 & 4.13\% & 0.0620 & 5.05\% & 0.0613 & 6.13\%  & 0.0653 & 0.0627 & 3.98\% & 0.0620 & 5.05\% & 0.0614 & 5.97\%\\ \hline
    \multirow{1}{*}{DRAGON}& 0.0662 & 0.0632 & 4.53\% & 0.0621 & 6.19\% & 0.0613 & 7.40\%  & 0.0662 & 0.0634 & 4.23\% & 0.0628 & 5.14\% & 0.0620 & 6.34\% \\
    +VI-MMRec-O& 0.0699 & 0.0672 & 3.86\% & 0.0665 & 4.86\% & 0.0655 & 6.29\% & 0.0699 & 0.0672 & 3.86\% & 0.0666 & 4.72\% & 0.0658 & 5.87\% \\
    +VI-MMRec-S& 0.0689 & 0.0662 & 3.92\% & 0.0655 & 4.93\% & 0.0646 & 6.24\% & 0.0689 & 0.0662 & 3.92\% & 0.0655 & 4.93\% & 0.0648 & 5.95\%\\ \hline
    \multirow{1}{*}{LGMRec}& 0.0647 & 0.0620 & 4.17\% & 0.0610 & 5.72\% & 0.0602 & 6.96\% & 0.0647 & 0.0620 & 4.17\% & 0.0610 & 5.72\% & 0.0602 & 6.96\% \\
    +VI-MMRec-O& 0.0681 & 0.0657 & 3.52\% & 0.0646 & 5.14\% & 0.0640 & 6.02\% & 0.0681 & 0.0654 & 3.96\% & 0.0647 & 4.99\% & 0.0640 & 6.02\% \\
    +VI-MMRec-S& 0.0671 & 0.0647 & 3.58\% & 0.0636 & 5.22\% & 0.0629 & 6.26\% & 0.0671 & 0.0645 & 3.87\% & 0.0638 & 4.92\% & 0.0630 & 6.11\%\\ \hline
    
    \multirow{1}{*}{DiffMM}& 0.0643 & 0.0615 & 4.35\% & 0.0606 & 5.75\% & 0.0598 & 7.00\% & 0.0643 & 0.0608 & 5.44\% & 0.0595 & 7.47\% & 0.0584 & 9.18\% \\
    +VI-MMRec-O& 0.0678 & 0.0653 & 3.69\% & 0.0644 & 5.01\% & 0.0637 & 6.05\% & 0.0678 & 0.0646 & 4.72\% & 0.0631 & 6.93\% & 0.0621 & 8.41\% \\
    +VI-MMRec-S& 0.0667 & 0.0642 & 3.75\% & 0.0633 & 5.10\% & 0.0627 & 6.00\% & 0.0667 & 0.0635 & 4.80\% & 0.0621 & 6.90\% & 0.0610 & 8.55\%\\ \hline
\end{tabular}
    }
    \vskip -0.1in
\end{table*}

\subsection{Hyper-parameter Analysis (RQ7)}
\label{subsec:RQ4}
We evaluate the influences of key hyper-parameters in VI-MMRec\footnote{The experimental results we report are conducted based on the VI-MMRec-O instance. Section~\ref{subsec:RQ1-2} demonstrates that VI-MMRec-O outperforms VI-MMRec-S in all cases.}. Due to space limitations, we report the Recall@10 for the two best-performing models (DRAGON and DiffMM) across all datasets. From the results in Figure~\ref{fig:hyper}, we have the following observations:

\noindent\textbf{Performance Comparison $w.r.t.$ $k$:} 
We conduct an in-depth analysis of the impact of the parameter $k$ on the construction of virtual user-item interaction matrices, as illustrated in Figure~\ref{fig:recall10_k}. The recommended value of $k$ is 10 for most datasets, with the exception of the TikTok dataset, where the optimal value is found to be 20. For most datasets, the top-20 most similar items tend to deviate from user preferences, thereby negatively affecting the inference of user preferences. In contrast, for the TikTok dataset, which incorporates more modalities, the top-20 items continue to contribute positively to the inference of user preferences.

\noindent\textbf{Performance Comparison $w.r.t.$ $\lambda$:} 
We empirically analyze the effect of the weight hyper-parameter $\lambda$ in our VI-MMRec. As shown in Figure~\ref{fig:recall10_lambda}, $\lambda = 1e^{-2}$ is recommended for most datasets, except for the TikTok dataset, where the optimal setting is $\lambda = 5e^{-3}$. We speculate that this difference is due to the TikTok dataset containing more modalities, requiring a reduction in the overall weight to ensure that the virtual user-item interactions do not outweigh the real user-item interactions.
\section{Conclusion}
In this paper, we aimed to mitigate the data sparsity problem in multimodal recommendation systems by constructing similarity-aware virtual user-item interactions to enrich sparse interaction signals. We propose VI-MMRec, a model-agnostic framework that augments real user-item interactions with high-confidence virtual user-item interactions derived from modality feature similarities. To ensure the quality of these virtual user-item interactions, we introduce two strategies-Overlay (independently aggregating modality-specific similarities) and Synergistic (holistically fusing cross-modal similarities). Moreover, we propose a statistically informed weight allocation mechanism that adaptively assigns weights to virtual user-item interactions based on dataset-specific modality relevance. As a plug-and-play framework, VI-MMRec seamlessly integrates with existing models to enhance their performance without modifying their core architecture. Its flexibility allows it to be easily incorporated into various existing models, maximizing performance with minimal implementation effort. Comprehensive experiments on six real-world datasets using seven advanced multimodal recommendation models demonstrated the effectiveness of our VI-MMRec.


\begin{acks}
This work was supported by the Hong Kong UGC General Research Fund no. 17203320 and 17209822, and the project grants from the HKU-SCF FinTech Academy.
\end{acks}

\balance
\bibliographystyle{ACM-Reference-Format}


\newpage
\appendix

\section{Appendix}
\subsection{Baseline}
We examine the performance of our VI-MMRec across seven advanced multimodal recommendation models:
\label{appendix:baseline}
\begin{itemize}[leftmargin=*]
\item \textbf{MMGCN} \cite{wei2019mmgcn} utilizes separate Graph Convolutional Networks (GCNs) for each data modality to capture modality-specific features, and then combines user-predicted ratings from all modalities to generate more accurate final predictions.
\item \textbf{DualGNN} \cite{wang2021dualgnn} incorporates a user-user graph to uncover latent preference patterns among users.
\item \textbf{SLMRec} \cite{tao2022self} adopts a self-supervised learning framework for multimodal recommendation by designing a node self-discrimination task, which reveals hidden multimodal patterns.
\item \textbf{FREEDOM} \cite{zhou2023tale} enhances LATTICE by freezing the item-item graph to maintain stable item relationships and reducing noise in the user-item graph, improving recommendation performance.
\item \textbf{DRAGON} \cite{zhou2023enhancing} exploits both heterogeneous and homogeneous graphs to learn high-quality representations for users and items.
\item \textbf{LGMRec} \cite{guo2024lgmrec} combines local embeddings, which capture fine-grained topological information, with global embeddings that consider hypergraph dependencies among items.
\item \textbf{DiffMM} \cite{jiang2024diffmm} introduces a carefully designed modality-aware graph diffusion model to enhance user representation learning by effectively capturing modality-specific information.
\end{itemize}

\subsection{Analysis of $O_{avg}$}
\label{appendix:math}
Specifically, $O_{real}^m=\frac{|R^m\cap R^+|}{|R^+|}$, where $R^m$ represents the set of virtual interactions constructed for each interaction $R^{+}$, with $k$ virtual interactions per record. Note that $|R^m| < k|R^+|$ due to overlaps in the virtual matrix.

Therefore, $O_{avg}=\frac{k|R^+|}{|\mathcal{U}||\mathcal{I}|}$, which approximates the probability of covering real interactions when $k|R^+|$ random points are placed in a matrix of size $|\mathcal{U}||\mathcal{I}|$, given there are $|R^+|$ fixed real interactions in the matrix. 

We further provide an approximate derivation process. For each real interaction, the probability of being selected in one random placement is $\frac{1}{|\mathcal{U}||\mathcal{I}|}$, and the probability of not being selected is $1-\frac{1}{|\mathcal{U}||\mathcal{I}|}$. Thus, the probability of not being selected in $k|R^+|$ placements is $(1-\frac{1}{|\mathcal{U}||\mathcal{I}|})^{k|R^+|}$. Further, the probability of being covered in $k|R^+|$ placements is: $p_{c}=1-(1-\frac{1}{|\mathcal{U}||\mathcal{I}|})^{k|R^+|}$.

Since there are $|R^+|$ real interactions, and each has the same (i.i.d.) probability of being covered, the expected number of covered real interactions is: $E_c=|R^+|p_c$. Thus, the expected coverage probability of real interactions is: $\frac{E_c}{|R^+|} = 1-(1-\frac{1}{|\mathcal{U}||\mathcal{I}|})^{k|R^+|}$. Given that $|\mathcal{U}||\mathcal{I}|$ is typically very large, we can approximate the expression as: $1-(1-\frac{1}{|\mathcal{U}||\mathcal{I}|})^{k|R^+|}\approx 1-\exp(-\frac{k|R^+|}{|\mathcal{U}||\mathcal{I}|})$.

Since recommendation datasets are highly sparse and $k$ is kept small (5,10,20), $\frac{k|R^+|}{|\mathcal{U}||\mathcal{I}|}$ is very small. Using the Taylor expansion of the exponential function: $\exp(-\frac{k|R^+|}{|\mathcal{U}||\mathcal{I}|})\approx 1-\frac{k|R^+|}{|\mathcal{U}||\mathcal{I}|}$. Thus, $1-\exp(-\frac{k|R^+|}{|\mathcal{U}||\mathcal{I}|})\approx 1-(1-\frac{k|R^+|}{|\mathcal{U}||\mathcal{I}|})=\frac{k|R^+|}{|\mathcal{U}||\mathcal{I}|}$.

We final calculate approximated $O_{avg}$ via $\frac{k|R^+|}{|\mathcal{U}||\mathcal{I}|}$.

\subsection{Related Work}
Recent works have increasingly focused on incorporating multimodal information to address the data sparsity problem in recommendation systems. A notable early contribution is VBPR \cite{he2016vbpr}, which integrates visual content as side information into matrix factorization \cite{rendle2009bpr} by leveraging item images to enhance recommendation performance. Building on this foundation, subsequent works \cite{chen2019personalized,liu2019user,yu2023multi,xu2024mentor} expand the scope by incorporating both visual and textual modalities to enrich item representations and improve recommendation performance. Advances in graph-based methods have further opened new opportunities for multimodal recommendation. For instance, MMGCN \cite{wei2019mmgcn} is the first to utilize Graph Convolutional Networks (GCNs) to extract modality-specific features from user-item interactions. To better capture commonalities in user preferences and item relationships, models such as DualGNN \cite{wang2021dualgnn} and LATTICE \cite{zhang2021mining} introduce user-user and item-item graphs, respectively. Building on LATTICE, FREEDOM \cite{zhou2023tale} enhances stability by freezing the item semantic graph and mitigating noise in the user-item bipartite graph. LOBSTER \cite{xu2025lobster} introduces global factors to capture shared features across all modalities.

Recently, self-supervised learning and inter-modal relationships have been explored to further improve recommendation systems. MMSSL \cite{wei2023multi} and MICRO \cite{zhang2022latent} adopt contrastive self-supervised learning to align multimodal signals with collaborative information, achieving improved performance without reliance on extensive labeled data. Additionally, BM3 \cite{zhou2023bootstrap} investigates inter-modal relationships, enhancing both recommendation accuracy and the quality of modality fusion. MENTOR \cite{xu2024mentor} proposes a tailored multi-level self-supervised learning task to align multiple modalities without losing user-item interaction information. Further innovations include leveraging complex graph-based structures and generative models. LGMRec \cite{guo2024lgmrec} employs hyper-graphs to capture intricate global and local relationships within multimodal information. DiffMM \cite{jiang2024diffmm} introduces diffusion models to reduce noise across modalities, significantly improving the robustness of multimodal recommendation systems. COHESION \cite{xu2025cohesion} unlocks the potential of composite graphs. FastMMRec \cite{xu2025best} takes the first step in applying graph propagation during the inference phase. HPMRec \cite{chen2025hypercomplex} leverages hyper-complex algebra to enhance multimodal information integration. These advancements collectively highlight the growing importance of multimodal and graph-based approaches in addressing challenges in recommendation systems.

Although the aforementioned multimodal recommendation works have demonstrated promising performance, their effectiveness remains constrained by the persistent challenge of data sparsity. Data sparsity arises because users typically interact with only a small subset of available items in real-world scenarios \cite{liu2024multimodal,xu2025survey,xu2025enhancing,xu2025mdvt}. As a result, many items that users might be interested in but lack interaction records are often arbitrarily treated as negative samples by existing models. To this end, we propose constructing virtual user-item interactions by leveraging the similarity of modality features among items that users have interacted with in real user-item interactions. Augmenting real user-item interactions through our virtual user-item interaction mitigates the limitations imposed by data sparsity, leading to a more accurate inference of user preferences.

\end{document}